\documentclass[12pt]{iopart}
\usepackage{setstack} 
\usepackage{graphicx,epstopdf} 
\usepackage{cite}

\begin{document}

\title[Efficient optical pumping and high optical depth in a HCPCF]{Efficient optical pumping and high optical depth in a hollow-core photonic-crystal fibre for a broadband quantum memory}

\author{Michael R Sprague$^1$, Duncan G England$^1$, Amir Abdolvand$^2$, Joshua Nunn$^1$, Xian-Min Jin$^1$, W Steven Kolthammer$^1$, Marco Barbieri$^{1}$, Bruno Rigal$^3$, Patrick S Michelberger$^1$, Tessa F M Champion$^1$, Philip StJ Russell$^{2}$, Ian A Walmsley$^{1}$  }

\address{$^1$ Clarendon Laboratory, Parks Rd, University of Oxford, OX1 3PU Oxford, UK}
\address{$^2$ Max Planck Institute for the Science of Light, Guenther-Scharowsky Str. 1, 91058 Erlangen, Germany}
\address{$^3$ \'{E}cole Polytechnique, Route de Saclay, 91120 Palaiseau, France}

\ead{michael.sprague@physics.ox.ac.uk}
\begin{abstract}
The generation of large multiphoton quantum states $-$ for applications in computing, metrology, and simulation $-$ requires a network of high-efficiency quantum memories capable of storing broadband pulses. Integrating these memories into a fibre offers a number of advantages towards realising this goal: strong light-matter coupling at low powers, simplified alignment, and compatibility with existing photonic architectures. Here, we introduce a large-core kagome-structured hollow-core fibre as a suitable platform for an integrated fibre-based quantum memory with a warm atomic vapour. We demonstrate, for the first time, efficient optical pumping   in a hollow-core photonic-crystal fibre with a warm atomic vapour, where 90~$\pm$~1$\%$ of atoms are prepared in the ground state. We measure high optical depths (3$\times$$10^{4}$) and, also, narrow homogeneous linewidths that do not exhibit significant transit-time broadening. Our results establish that kagome fibres are suitable for implementing a broadband, room-temperature quantum memory. 

\end{abstract}

\maketitle

\section{Introduction}

Quantum memories are critical for the synchronisation of probabilistic components in large-scale quantum networks \cite{Sangouard2011a} and have been implemented using a range of protocols and physical systems \cite{Julsgaard2004, NatureChoi2008, Zhao2009, NatureHedges2010, Specht2011, NatureHosseini2011, Clausen2011,Zhang2011, Lee2011, Lee.2012}. In the context of increasing the rate of multiphoton generation from probabilistic single-photon sources, a key metric is the product of the memory efficiency $\eta$ and the time-bandwidth product (TBP) $B$, the ratio of the storage time to the pulse duration \cite{Nunn2012}.  Recently, we demonstrated, using a warm atomic vapour of cesium, a far-off-resonant Raman memory capable of storing broadband (1.5 GHz), single-photon-level pulses \cite{Reim.2010, Reim2011, Reim2012} with a TBP $B$~=~3000 and an efficiency $\eta$~=~ 30$\%$, which would yield, for instance, a thousand-fold enhancement in the rate of generation of two synchronised single photons.

Towards this goal, a quantum memory in an integrated architecture is attractive for several reasons. First, guided-wave optics can confine an optical mode to a small area over a distance longer than is possible with diffractive optics. The optical power needed to achieve strong light-matter coupling can therefore be reduced; in practice this could increase the attainable memory efficiency \cite{Reim.2010}. Second, an integrated platform can be easily interfaced with existing photonic architectures \cite{Schreiber2012, Shadbolt2012, Metcalf2012} as well as easily scaled-up to build multiple units.  Third, an integrated system simplifies the spatial overlap of the various beams involved in the memory interaction.  One possible approach is to use on-chip waveguides coupled to rare-earth doped solids cooled to several Kelvin \cite{Saglamyurek2011} or miniature vapour cells \cite {Wu2010}. An especially interesting approach, however, is to use hollow-core photonic crystal fibres, which have lower transmission losses than chip-based waveguides and have an empty core that can be filled, at room temperature, with an atomic vapour such as cesium. 

\section{Kagome-structured hollow-core fibre}

Hollow-core photonic crystal fibres are divided into two main types: photonic bandgap fibres and kagome-structured photonic-crystal fibres. Photonic bandgap fibres (PBGFs) possess a true bandgap as their guidance mechanism, which prevents light from propagating in the fibre cladding and confines the light within the core \cite{Cregan1999}. Electromagnetically induced transparency \cite{Ghosh2006, Bajcsy2009a}, four-wave mixing \cite{Londero2009}, and two-photon absorption \cite{PhysRevA.83.033833} have been previously demonstrated with alkali atoms in PBGFs with a 6-$\mu$m diameter core. The cladding of kagome-structured photonic crystal fibres (see figure 1(b)) does not possess a photonic bandgap, but guides by a mechanism that, while still being elucidated \cite{Benabid2002,Couny2006, Pearce2007}, appears to be related to anti-resonant reflections \cite{Benabid2011}. Kagome fibres can easily be made with large core diameters (25 to 30 $\mu$m) and benefit from spectral guidance over several hundred nanometers at the cost of slightly higher transmission losses than PBGFs \cite{Pearce2007, Light2007}. A large core is important to facilitate loading atoms, as will be discussed in section~3, and to allow for sufficient time to store and retrieve a pulse of light, as will be discussed in section 4.

To test its suitability for a quantum memory, we mounted a 20-cm length of kagome fibre in an ultra-high-vacuum system (base pressure 5 x $10^{-10}$ Torr) with two viewports to provide optical access for the coupling lenses mounted outside the vacuum system (see figure~1(a)). The core diameter was 26 $\mu$m, shown in figure~1(b), and the transmission loss of the fibre at a wavelength of 852 nm was measured to be 1 dB/m. A glass ampoule, broken under vacuum using a flexible bellows, provided a source of cesium atoms for loading the fibre, as shown in figure~1(a). The coupling efficiency of the optical modes into the fibre was 65$\%$. The collimated output of the kagome fibre coupled into a single-mode fibre with 67$\%$ efficiency, indicating the output was, to good approximation, single mode (see also figure 1(c)). For use as a quantum memory, it is essential to observe a high optical depth and efficient optical pumping in the kagome fibre, as will be detailed in the following sections.

\begin{figure}
\center \includegraphics[scale=0.45]{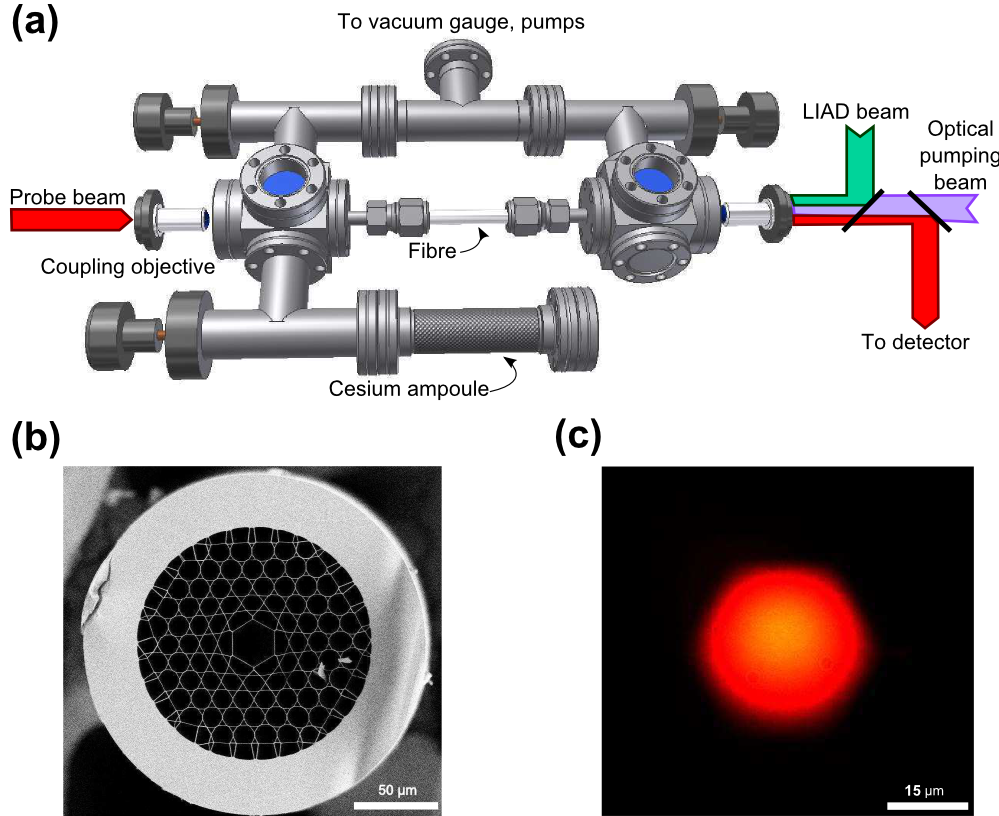}
\caption{(a) Schematic of the high-vacuum system showing the ensheathed fibre, location of the cesium ampoule and the orientation of optical beams. The two viewports on the top of the apparatus were to monitor the amount of cesium in the vacuum chamber. The LIAD and optical pumping beam are combined with a spectral filter. (b) Scanning electron microscopy image of the kagome-structured hollow-core fibre with core diameter of 26~$\mu$m and pitch 13~$\mu$m. (c) Optical image of the fundamental mode at the face of the fibre.}
\end{figure}

\section{Optical depth}

The optical depth $d$, defined as the on-resonance absorption $d=-\ln\left(I/I_{0}\right)$ with no inhomogeneous broadening, quantifies the strength of the light-matter coupling in an ensemble-based memory. It is the parameter that defines the efficiency of the memory interaction for an equivalence-class of memory protocols \cite{Gorshkov2007, Gorshkov2007a}. For a Raman memory, the coupling $C^{2}$ is given by $C^{2} \approx d \left(\frac{\gamma}{\delta} \right) \left( \frac{\Omega}{\Delta} \right)^{2}$, where $\gamma$ is the homogeneous linewidth, $\delta$ is the bandwidth of the pulse to be stored, $\Omega$ is the Rabi frequency of the control field and $\Delta$ is the excited-state detuning \cite{Nunn2007}. Efficient memory operation requires $C^{2} \geq 1$. In the adiabatic regime, the Rabi frequency is much less than the detuning and, therefore, the condition for efficient memory becomes $d \gg \left(\frac{\delta}{\gamma}\right)$. In the presence of inhomogeneous broadening, the effective optical depth $d^{*}$ is reduced by the ratio of the inhomogeneous $\gamma_{i}$ to the homogeneous broadening, giving the condition $d^{*} \gg \left(\frac{\delta}{\gamma_{i}}\right)$. As an example, storing 1.5~GHz pulses in cesium at room temperature (cf. \cite{Reim.2010, Reim2011}) requires an effective optical depth $d^{*} \gg 4$.

The optical depth in the fibre was measured with a low-power, continuous-wave probe beam (10 nW) swept over the Doppler-broadened resonance. An avalanche photodiode detected the transmitted light. The normalised transmission spectrum for the ground-state hyperfine manifold $F$ was fit to the function $T_{F}(f)=\exp\left(-d^{*} \cdot {\sum}S_{FF'}\exp\left(\frac{-\left(f-f_{F'}\right)^{2}}{2\sigma^{2}}\right)\right)$, where the summation is over the excited-state hyperfine manifolds $F'$.  The Doppler width $\sigma$ and the optical depth were free parameters and the relative transition strengths of the hyperfine manifolds $S_{FF'}$ and the frequency spacing ($f_{F'}-f_{F''})$ of the excited-state hyperfine manifolds were fixed parameters. Simultaneously, we performed saturation absorption spectroscopy in a bulk vapour-cell in order to calibrate the frequency span of the sweep. The optical depth in the fibre was $d^{*}$~=~5.7~$\pm$~0.1 for the $6^{2}S_{1/2}(F=3)$ to $6^{2}P_{3/2}$ transition, as shown in figure 2(a), and $d^{*}$~=~5.7~$\pm$~0.3 for the $6^{2}S_{1/2}(F=4)$ to $6^{2}P_{3/2}$ transition. We estimated that 85$\%$ of these atoms were confined to the core of the fibre by measuring the optical depth between the viewport and the fibre face (spaced by 3~mm). The optical depth is an order of magnitude larger than obtained previous measurements in a PBGF \cite{Ghosh2006, Light2007}, and approaches the level needed for an efficient, broadband quantum memory in cesium.

\begin{figure}
\center \includegraphics[scale=0.59]{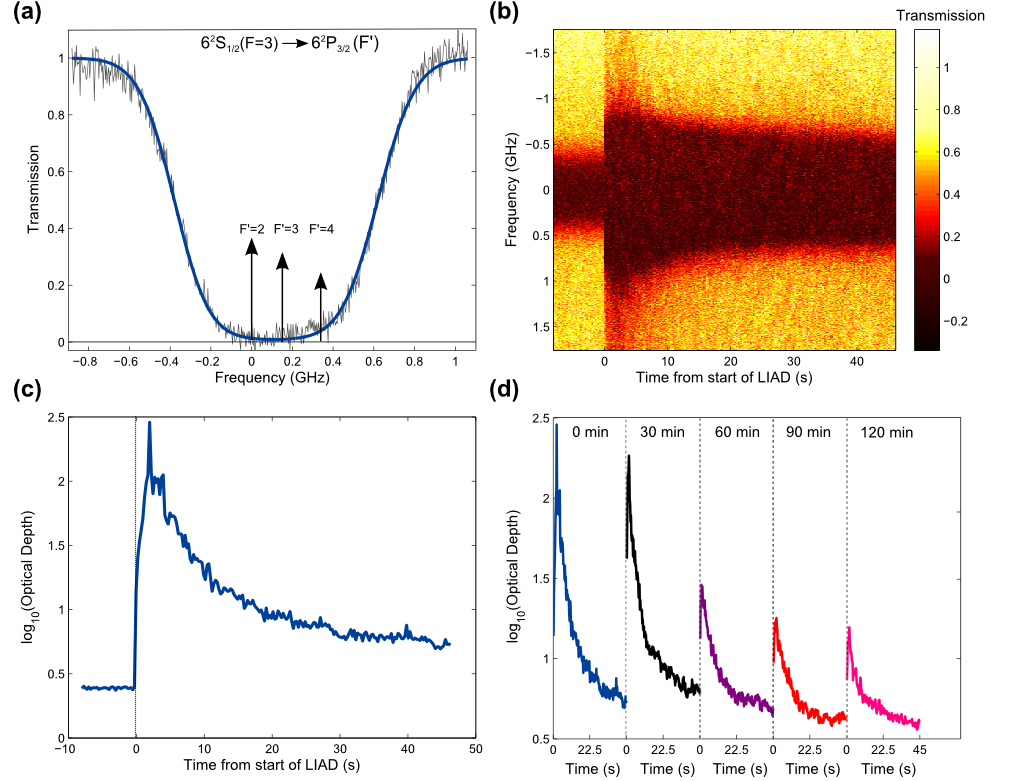}
\caption{(a) Transmission spectrum through the kagome fibre \emph{without} LIAD, and heated to 363~K, showing measurement (thin black line) and fit (thick blue line). Arrows indicate location of hyperfine manifolds with the lengths are scaled relative to the transition strength. (b) Transmission spectra during the application of LIAD (80 mW beam at 780 nm wavelength). (c) Extracted effective optical depth $d^{*}$ from figure 2(b), plotted as the logarithm base ten. The dashed line indicates the onset of LIAD. (d) Repeated application of the same 45-s LIAD pulse at 30-minute intervals for 120 minutes.}
\end{figure}

\subsection{Light-induced atomic desorption}       

The optical depth can be further increased through light-induced atomic desorption (LIAD), a transient process where alkali atoms are ejected from a surface by interaction with light far-detuned from resonance \cite{Meucci1994}. This effect was previously observed with rubidium in uncoated PBGFs \cite{Slepkov2008, Bhagwat2009}. Here, we study the evolution of the optical depth during repeated exposure to LIAD. The LIAD beam was derived from a 780-nm, continuous-wave diode laser amplified using a tapered amplifier, and we coupled 80~mW, counter-propagating, through the uncoated kagome fibre (see figure 1(a)). After application of the LIAD beam, the optical depth rapidly increased within 100~ms by two orders of magnitude and then slowly decreased over one minute, even though the LIAD beam remained on. Figure 2(b) shows an example of how the transmission spectra changed as LIAD was applied. The effective optical depth is plotted in figure 2(c), and reached an estimated maximum $d^{*}$ = 300 $\pm$ 50. The fit Doppler width $\sigma$ during LIAD corresponded to a temperature of 480 K (compared to the cell temperature of 363 K), although in this transient regime atoms may not have reached thermal equilibrium with the fibre \cite{Slepkov2008}. We repeated the application of LIAD at 30-min intervals for two hours, as shown in figure 2(d), and observed a third time-scale; the peak optical depth decreased with each application of LIAD. Further measurements showed that a full recovery of the maximum optical depth required several hours of quiescence. This is qualitatively consistent with previous observations in a PBGF, where a phenomenological model was proposed involving the formation and plasmon-excitation of nano-clusters \cite{Slepkov2008, Bhagwat2009}. To summarise this section, we have shown that an extremely high optical depths, $d^{*}$~=~300 and $d$~=~3 $\times 10^{4}$ can be reached with LIAD in kagome fibres, a necessary condition for an efficient quantum memory.

\section{State preparation}

\subsection{Optical pumping}
In a $\Lambda$-system quantum memory, the atoms must be efficiently prepared into the ground state by means of optical pumping. At room temperature, however, the two hyperfine ground-state manifolds of cesium are essentially equally populated; thermal excitations in the storage state can be read-out during retrieval of the stored excitation via spontaneous Raman scattering, contributing noise to the memory process and reducing the fidelity of the memory. In addition, in the limit of equal populations, the competing processes of absorption and emission reduce the memory efficiency; therefore, efficient optical pumping is critical to realise a quantum memory.  In a confined geometry, such as in a hollow-core fibre, the atoms must be optically pumped during the time taken for the atoms to traverse the optical mode in the fibre. Otherwise, collisions with the fibre wall may destroy the spin-polarisation of the atoms. The limiting time-scale for optical pumping is the excited-state lifetime, since multiple spontaneous decay events are required to prepare all the atoms in the ground state. The transit time must, then, exceed the excited-state lifetime. In a 26-$\mu$m fibre, the mean transit time across the optical mode (with a 12~$\mu$m Gaussian beam width) is $\sim$100 ns and the excited-state lifetime of the $D_{2}$ line is 30 ns \cite{SteckCs}. This motivated an experimental study of the maximum optical pumping efficiency in the fibre.

We measured the optical pumping efficiency as a function of optical pumping power, parameterised by the Rabi frequency $\Omega_{p}$~=~$\frac{\vec{d}\cdot\vec{E}}{h}$, where $\vec{d}$ is the dipole moment and $\vec{E}$ is the electric field of the pump beam. A weak probe beam (10~nW) was scanned in frequency over the $6^{2}S_{1/2}(F=3)$ to $6^{2}P_{3/2}$ hyperfine manifold in order to measure the change in optical depth due to the counter-propagating optical pumping beam, derived from a separate diode laser. The frequency of the optical-pump laser was locked to a sub-Doppler peak and the linewidth was 1~MHz. Figure~3(a) shows the change in the transmission over the entire Doppler-broadened profile due to optical pumping, where the Rabi frequency of the pump is 650~MHz. From the change in optical depth, we extracted the optical pumping efficiency, that is the percentage of atoms in the ground state, which is plotted as a function of the Rabi frequency in figure~3(b). The maximum pumping efficiency we measured was (90 $\pm$ 1)$\%$ for $\Omega_{p}$ =~700~MHz, which exceeded the Doppler-broadened linewidth of the transition (full-width half-maximum~=~420 MHz). Figure 3(b) represents the first observation of efficient pumping in a hollow-core fibre with a warm alkali vapour; previous experiments with small-core PBGFs were limited to 55$\%$ optical pumping efficiency \cite{Londero2009}. Our results imply that the narrow-linewidth pump beam is sufficiently power-broadened to pump all velocity classes in the Doppler-broadened transition. This represents a significant advantage of the hollow-core geometry as compared to free-space systems: high Rabi frequencies can be achieved with relatively low, and easily accessible, powers ($<$1~mW). This is in contrast to previous experiments with bulk vapour cells \cite{Reim.2010, Reim2011}, where, instead of power broadening, the optical pumping required collisions with a buffer gas to sweep all the atoms through the velocity class resonant with the frequency of the laser diode.  

\begin{figure}
\center \includegraphics[scale=0.59]{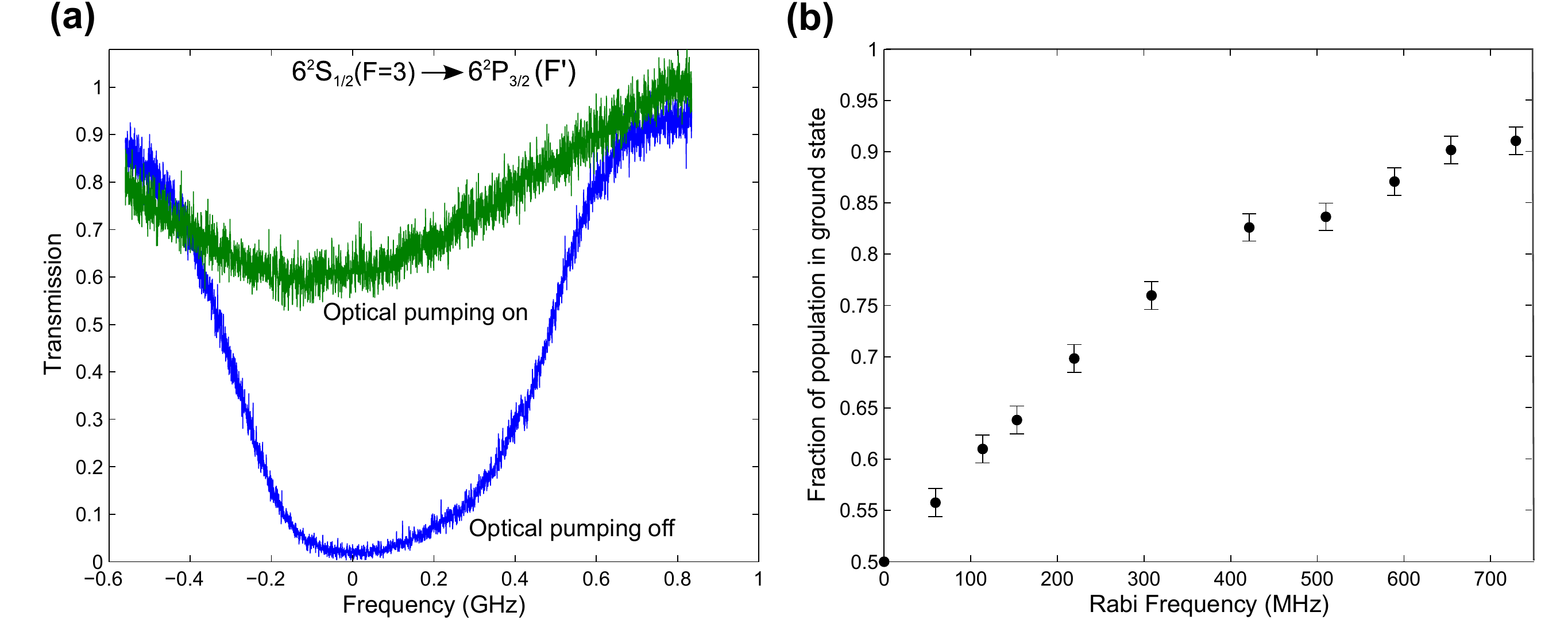}
\caption{(a) Optical pumping of the Doppler-broadened transition with a fixed-frequency pump at a Rabi frequency of 650 MHz.  (b) The fraction of the population pumped into the ground hyperfine state ($F=4$) as a function of the Rabi frequency of the optical pumping beam.}
\end{figure}

\subsection{Saturated absorption spectroscopy}

Next, we examined the linewidth of the atoms confined within the core of the fibre. If the transit time were comparable to the excited-state lifetime, the homogeneous linewidth would be broadened to a width greater than the natural linewidth of the transition. We directly measured, with saturation absorption spectroscopy, the homogeneous linewidth from atoms confined within the core of the fibre. To resolve the excited-state hyperfine manifold, a weak probe beam (10 nW), coupled into the fibre, was scanned over the Doppler-broadened transition and a higher-power, counter-propagating pump beam, derived from the same laser, was launched into the fibre. The resulting spectrum is shown in figure~4(a). The width of the $6^{2}S_{1/2}(F=3)$  to $6^{2}P_{3/2}(F'=4)$ transition was measured as a function of the pump intensity $I$ scaled to the experimentally determined saturation intensity $I_{\mathrm{sat}}$. Outside the vacuum chamber, saturated absorption spectroscopy was performed in a bulk vapour-cell to estimate the homogeneous linewidth in a parameter range when transit-time broadening is known to be negligible; this served as a verification to our analysis. The Doppler-free linewidths are plotted in figure 4(b), along with a weighted nonlinear least-squared fit according to the equation $\Gamma = \Gamma_{0} \sqrt{1 + \frac{I}{I_{\mathrm{sat}}}}$, where $\Gamma$ is the power-broadened homogeneous linewidth and $\Gamma_{0}$ is the homogeneous linewidth. The homogeneous linewidth in the fibre was 6~$\pm$~2~MHz and the saturation power was 50~$\pm$~20 nW, which agrees with the homogeneous linewidth in the vapour cell of 6.5~$\pm$~0.6~MHz; both are consistent with the natural linewidth of 5.2~MHz \cite{SteckCs}. There was no apparent transit-time broadening within the resolution of our measurement. In contrast, previous measurements done with commercially available hollow-core photonic-crystal fibre, with a 6 $\mu$m core, measured a homogeneous linewidth of 27~$\pm$~3 MHz \cite{Slepkov2010}, indicating significant transit-time broadening. Furthermore, the saturation spectroscopy supports the optical pumping measurements in section 3, both of which show that the the transit time exceeds the excited state lifetime in the large-core kagome fibre, a critical condition for initialising pure quantum states. 

\begin{figure}
\center \includegraphics[scale=0.59]{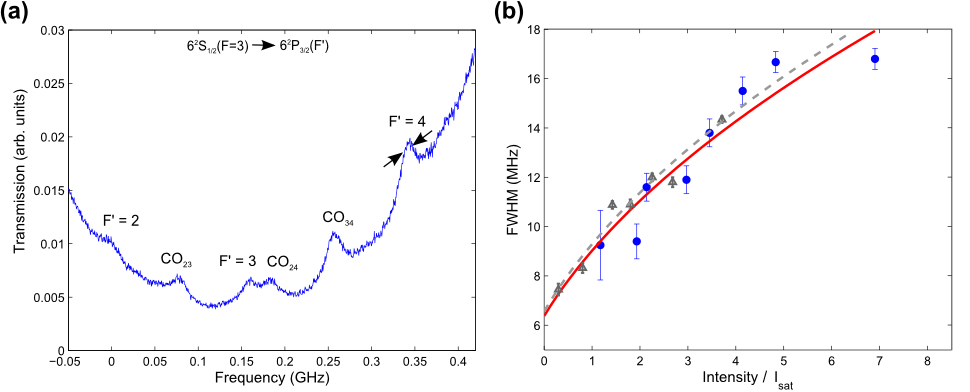}
\caption{(a) Saturation absorption spectroscopy of atoms inside the kagome fibre. The probe power was 10~nW (0.2$\cdot I_{\mathrm{sat}}$). $\mathrm{CO_{23}}$ refers to the cross-over resonance of the $F'=2$ and $F'=3$ excited-state hyperfine manifold.  (b) Width of the ($F=3$) to ($F'=4$) transition as a function of intensity (power). Blue circles refer to measurements in the kagome fibre, and the red solid line is the fit (see text). Grey triangles refer to measurements made in the bulk vapour cell, and the grey dashed line is the fit. }
\end{figure}

\section{Conclusion and future directions}

We have presented strong evidence that kagome-structured hollow-core fibres loaded with cesium vapour are suitable for implementing a quantum memory. We have observed, for the first time, efficient optical pumping in a hollow-core photonic crystal fibre, measured the narrowest homogeneous linewidth in such a system and demonstrated an extremely high optical depth. These represent key parameters for the initialisation of a Raman quantum memory. The initial memory lifetime will be limited by the transit time of atoms within the fibre-confined mode ($\sim$100 ns). However, with a 300 ps pulse \cite{Reim.2010, Reim2011}, this represents a TBP of several hundred, making it already useful for enhancing the rate of multiphoton generation  \cite{Nunn2012}.  Further, the spin-coherence lifetime could potentially be extended by coating the fibre walls with a spin-preserving compound \cite{Ghosh2006, Light2007}, which may be especially promising for large-diameter hollow-core fibres such as kagome fibres. In addition, splicing the kagome fibre to a solid-core, single-mode fibre under vacuum \cite{Benabid2005} would enable the realisation of an integrated, `plug-and-play' quantum memory. 

\ack
The authors thank Mike Tacon and his team at the departmental machine shop for their assistance, and Peter Mosley for his helpful input. This work was supported by the Royal Society, EU IP Q-ESSENCE (248095), EU ITN FASTQUAST (PITN-GA-2008-214962), EPSRC (EP/J000051/1), AFOSR EOARD (FA8655-09-1-3020), EU Marie-Curie Fellowship (PIIF-GA-2011-300820, XMJ), the Clarendon Fund (MRS) and St. Edmund Hall (MRS).  

\section{References}
\bibliographystyle{nature}
\bibliography{HCF}

\end{document}